\documentclass[12pt,preprint]{aastex}

\usepackage{multirow}

\newcommand{\Msun}{~M_\odot}

\newcommand{\kms}{\rm ~km~s^{-1}}
\newcommand{\ergs}{\rm ~erg~s^{-1}}

\newcommand{\ml}{~\Msun ~\rm yr^{-1}}

\shorttitle{X-ray evolution of SN 2010jl}
\shortauthors{Chandra et al.}

\begin{document}

\title{STRONG EVOLUTION OF X-RAY ABSORPTION IN THE TYPE IIn SUPERNOVA SN 2010jl}

\author{Poonam\,Chandra\altaffilmark{1},
Roger\,A.\,Chevalier\altaffilmark{2},
Christopher\,M.\,Irwin\altaffilmark{2},
Nikolai\,Chugai\altaffilmark{3},
Claes\,Fransson\altaffilmark{4}, and
Alicia\,M.\,Soderberg\altaffilmark{5}}

\altaffiltext{1}{Department of Physics, Royal Military College of
  Canada, Kingston, ON, K7K 7B4, Canada (Poonam.Chandra@rmc.ca)}
\altaffiltext{2}{Department of Astronomy, University of Virginia, P.O. Box 400325,
Charlottesville, VA 22904-4325}
\altaffiltext{3}{Institute of Astronomy of Russian Academy of Sciences, Pyatnitskaya St. 48,
109017 Moscow, Russia}
\altaffiltext{4}{Department of Astronomy, Stockholm University, AlbaNova, SE-106 91 Stockholm,
Sweden}
\altaffiltext{5}{Smithsonian Astrophysical Observatory, 60 Garden St., MS-20, Cambridge, 
MA 02138}

\begin{abstract}

We report two epochs of {\em Chandra}-ACIS
X-ray imaging spectroscopy  of
the nearby bright Type IIn supernova SN 2010jl,
taken around 2 months and then a year after the explosion. 
The majority of the X-ray emission in both the spectra 
is characterized by a high temperature ($\ga 10$ keV) and is likely to be
from the forward shocked region resulting from circumstellar interaction.
The absorption column density in the first spectrum is  high ($\sim 10^{24}$ cm$^{-2}$), more than 3 orders of magnitude 
higher than the Galactic absorption column, and we attribute it to absorption
by  
circumstellar matter.   In the second epoch observation, 
the column density has decreased by a factor of 3, as expected for shock propagation
in the circumstellar medium. 
The unabsorbed $0.2-10$ keV luminosity at both epochs is $\sim 7\times 10^{41}\ergs$.
The 6.4~keV Fe line clearly present 
in the first spectrum is not detected in the second spectrum.
The  strength of the fluorescent line is roughly that expected for the column density of
circumstellar gas, provided the Fe is not highly ionized.
There is also  evidence for an absorbed power law component
in both the spectra, which we attribute to a background ultraluminous X-ray source.   

\end{abstract}

\keywords{Supernovae: General---Supernovae: Individual (SN 2010jl)---hydrodynamics---
circumstellar matter---X-rays: general}

\section{Introduction}

Supernova (SN) 2010jl was discovered on 2010
Nov 3.5 (UT) at a magnitude of 13.5 \citep{np10}, and
brightened to magnitude 12.9 over the next day,
showing that it was discovered at an
early phase.
Pre-discovery observations indicate an explosion date in early October 2010
\citep{stoll11}.
Spectra on 2010 Nov 5 show that it is a Type IIn event
\citep{benetti10}. 
The apparent
magnitude is the brightest for a Type IIn SN since SN 1998S.
SN 2010jl is associated with the galaxy UGC 
5189A at a distance of 50 Mpc ($z=0.011$),
implying that SN 2010jl belongs to the class of luminous SNe IIn with an
absolute magnitude  $<-20$.
{\em Hubble Space Telescope} (HST) images of the site of the SN taken 
a decade before the SN
indicate that,
unless there is a chance coincidence of a bright star with the SN site,
the progenitor star had an initial mass  $\ga 30~M_{\odot}$
\citep{smith10}.
Optical spectra  give evidence for a dense circumstellar medium (CSM) expanding 
around the progenitor star with speeds of $40-120\kms$ \citep{smith10}.
\citet{stoll11} found that the host galaxy
is of low metallicity, supporting
the emerging trend that luminous  SNe
occur in low metallicity environments. They determine the metallicity $Z$ of the
SN region to be $\lesssim0.3~  Z_\odot$.

{\em Spitzer} observations showed
a significant infrared (IR) excess in SN 2010jl, indicating either
new dust formation or the heating of circumstellar  dust in an 
IR echo \citep{andrews11}. 
\citet{andrews11}  attributed the IR excess to
pre-existing dust and inferred a  massive CSM around SN
2010jl, possibly suggesting a luminous blue variable (LBV) like progenitor. 
\citet{smith12}  found signatures of new dust formation in the
post-shock shell of SN 2010jl from their
multi-epoch spectra. While a significant IR excess is present, 
the SN 
does not show large reddening, indicating that the dust
does not have a spherically symmetric distribution about the SN
  \citep{andrews11}.
The {\em Swift} on-board X-ray Telescope (XRT) detected X-rays 
from SN 2010jl on 2010 Nov $5.0-5.8$ \citep{immler10}.
 Assuming a temperature of 10 keV and a Galactic absorption column of
$N_H=3.0\times 10^{20}$ cm$^{-2}$, \cite{immler10}
obtained an unabsorbed
X-ray luminosity of $3.6\pm 0.5\times 10^{40}$ erg s$^{-1}$
in the $0.2-10$ keV band.

After the detection of SN 2010jl with the {\em Swift}-XRT, we
triggered {\em Chandra} observations of the SN at two epochs, in Dec 2010
and Oct 2011,
and we present the results here (Section~\ref{observation}). 
We discuss the 
significant changes in the two {\em Chandra} observations taken 10 
months apart in Section~\ref{discussion}. 

\section{Observations and Analysis}
\label{observation}

\subsection{Observations}

\begin{figure}
\centering
\includegraphics[angle=-90,width=0.90\textwidth]{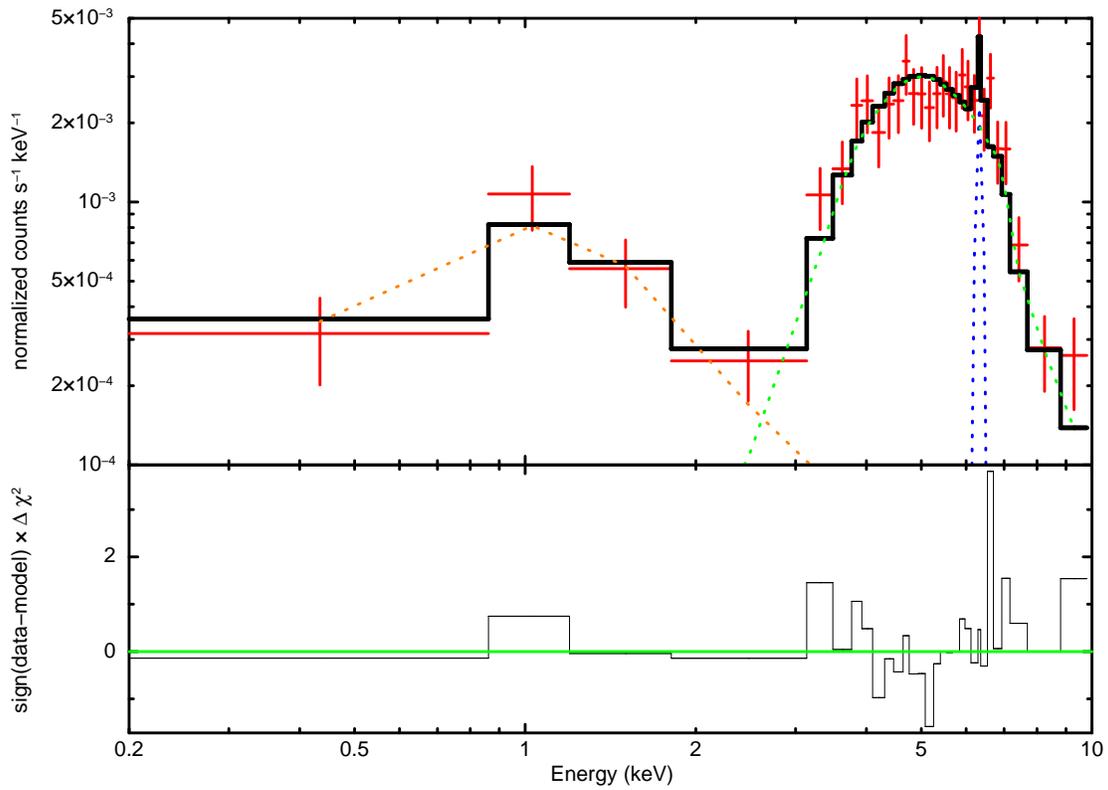}
\caption{ Best fit {\em Chandra} spectrum of SN 2010jl taken in Dec 2010.
The spectrum is best fit with a high-$T$, high-$N_H$ thermal component,
a Gaussian at 6.4 keV and a power law with photon index $\Gamma=1.76$.
}
\label{first}
\end{figure}

\begin{figure}[h]
\centering
\includegraphics[angle=-90,width=0.90\textwidth]{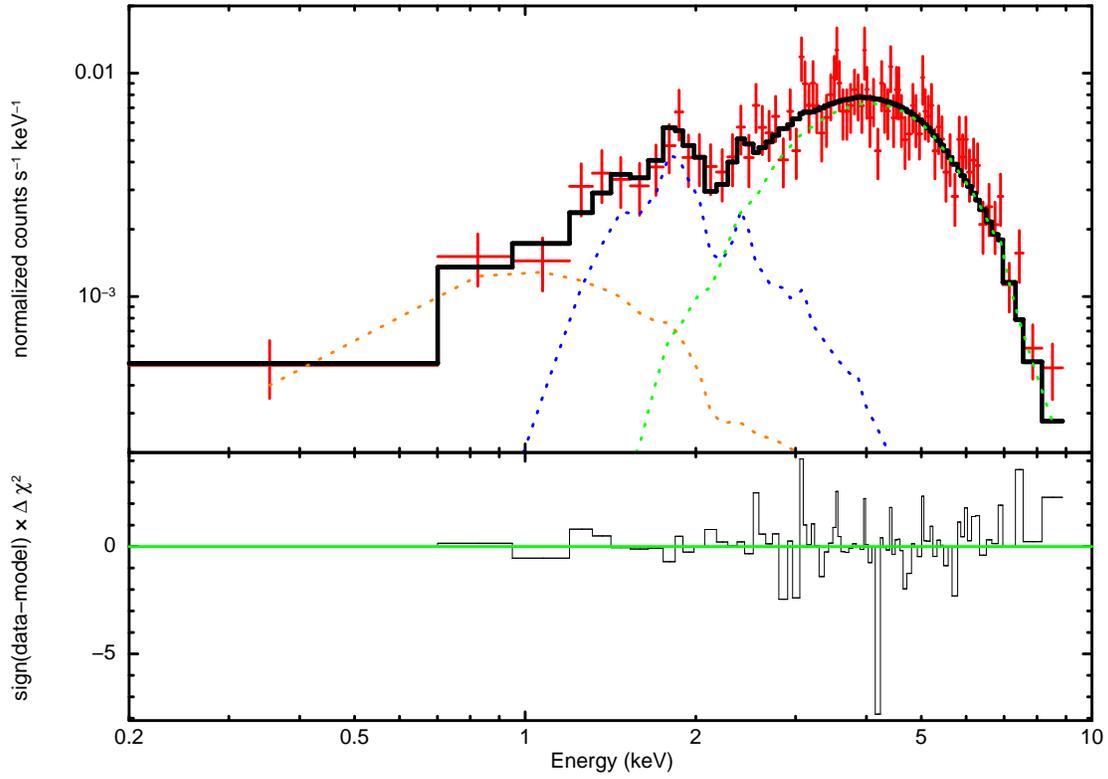}
\caption{Best fit {\em Chandra} spectrum of SN 2010jl taken in October
2011. The 
plot is for the model when the column densities of both temperature components (green 
dashed line and the blue dashed line) are allowed to vary independently in the
fit.} 
\label{second}
\end{figure}

The {\em Swift} detection of SN 2010jl allowed us to trigger
 our approved {\em Chandra}  Cycle 11
program in Dec 2010. We again observed SN 2010jl
in Oct 2011 under Cycle 13 of {\em Chandra}.
The first observation (Figure \ref{first})  took place under the proposal 
\# 11500430 starting 2010 Dec 7 at 04:22:53 Hrs (UT) for an
exposure of 19.05 ks and then on 2010 Dec 8 at 00:50:20 Hrs (UT)
for a 21.05 ks exposure. The observations were taken with the
ACIS-S without grating in a VFAINT mode. A total of 39.58 ks exposure time was
used in the data analysis and 468 counts were obtained with
a count rate of
$(1.13\pm0.05)\times10^{-2}$ cts s$^{-1}$. The second set of observations (Figure \ref{second})
took place under our proposal \# 13500593 starting on 2011 Oct 17 at
20:25:09 Hrs (UT) for an exposure of 41.04 ks. The observations were
again taken with the ACIS-S in the VFAINT mode with grating NONE. 
In a total of 40.51 ks usable exposure time, we obtained 1342 total counts,
i.e. a count rate of $(3.29\pm0.09)\times10^{-2}$ cts s$^{-1}$. 
We extracted the spectra using CIAO 
software\footnote{\url{http://asc.harvard.edu/ciao/}} and
used HEAsoft\footnote{\url{http://heasarc.gsfc.nasa.gov/docs/software/lheasoft/}} 
to carry out the spectral analysis.

\subsection{Spectral Analysis}
\label{spectralanalysis}

As can be seen in Figures \ref{first} and \ref{second},  the normalized count rate is  higher in the 
Oct 2011 spectra.   This does not necessarily indicate 
higher intrinsic emission from the SN at the later time, because the count rates
are absorbed count rates and  the unabsorbed emission  
depends on  the intervening column density, which may change. 
In the $5.5-7.5$ keV
range, the fluorescent 6.4 keV Fe line is present in the first spectrum
but not the second (Figure~\ref{comparison}).
Here we  carry out a detailed spectral analysis 
of both spectra and determine the 
significance of various emission components along with the Fe 6.4~keV line.

\subsubsection{December 2010 spectrum}
\label{dec2010}

For the December 2010 spectrum, we first fit the high temperature component in the $2-10$~keV 
energy range, where most of the emission in the {\em Chandra}
energy band lies. We fit the spectrum
in this range with an absorbed Mekal model \citep{mewe85,liedahl95} with 
metallicity $Z=0.3~ Z_{\odot}$ \citep{stoll11}.
The preferred Mekal temperature always hit the upper bound of the temperature allowed
in the Mekal model, i.e.\ 79.9 keV. The column density  is also very high, with
$N_H\approx 10^{24}$ cm$^{-2}$. 
Since the
temperature in the best fit models seems  high, we checked for the possibility of
 non-thermal X-ray emission and fit a power law model. The column
density in this fit is consistent with that of the Mekal model; however the 
photon index is too small to be physically plausible ($\Gamma=0.33$), and we disfavor
the non-thermal model. 

\begin{figure}[t]
\centering
\includegraphics[angle=-90,width=0.92\textwidth]{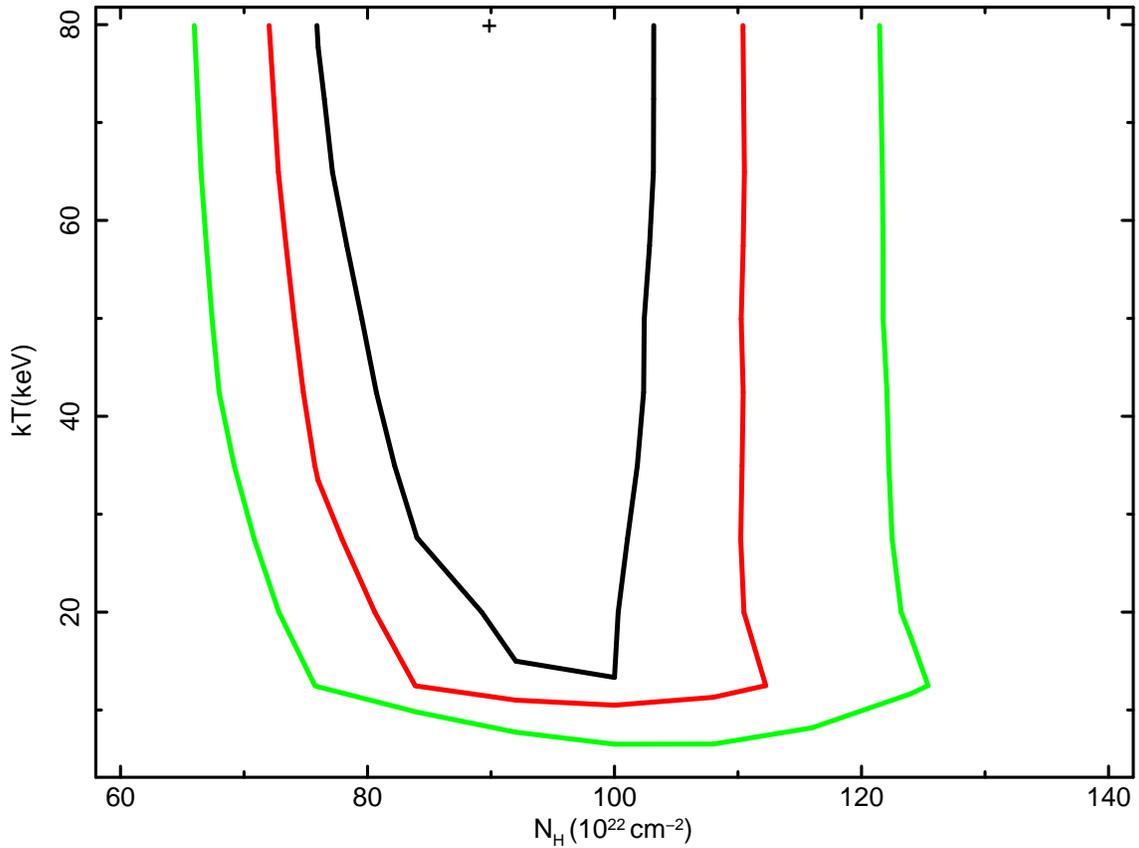}
\caption{The plot shows 68\% (black), 90\% (red) and
99\% (green) confidence contours for the $N_H$ 
versus $kT$ confidence contours plot for the December 2010 spectrum. The column density
is well constrained while the temperature is unconstrained on the high side.
}
\label{confidence1}
\end{figure}

When we  plot the
confidence contours of $N_H$ versus $kT$, 
the   column density in our fits 
is well constrained, but the upper bound of the temperature is
not  constrained 
(Figure~\ref{confidence1}). 
We established a lower bound on the temperature by assuming a value
and finding the goodness of fit;
 $T= 8$ keV gives a good fit with acceptable $\chi^2$ value,
but not lower values. Thus 8~keV is a lower limit on the temperature of the
main X-ray emission component of the SN.
For the lower temperature component in the $0.2-2$ keV
range,  we fix the absorption column to the Galactic value of
$3.0\times10^{20}$ cm$^{-2}$ since the absorption column
for this component is very poorly constrained.
This component is best fit with a temperature of $\sim 2$~keV or a 
power law with $\Gamma= 1.76$.

Figure \ref{first} shows the best fit to the
{\em Chandra} spectrum taken between 
2010 Dec 07.18 UT and 2010 Dec 08.03 UT, and the
best fits are tabulated in Table~\ref{tab:xspec}. 
Our complete model is the Absorption$\times$Powerlaw $+$
Absorption$\times$(Mekal$+$Gaussian).
97.9\% of the total unabsorbed flux
in the {\em Chandra} band is carried by the high temperature component.
The Fe 6.4 keV line carries 1.7\% of the total flux while the low-$N_H$ component
has only 0.4\% of the total flux.
The rest energy of the Fe line is $6.39\pm 0.06$ keV, which is consistent with the K$\alpha$ line.
The equivalent width of the line is $EW_{Fe}=0.2\pm 0.1$ keV.

\subsubsection{October 2011 Spectrum}

The {\em Chandra} spectrum taken between 2011 Oct $17.85-18.33$~UT is
 different from the first epoch spectrum.
We first fit the high temperature component between $2-10$ keV. The temperature in
this case also reaches the Mekal model upper limit of 79.9 keV.
To consider the possibility of non-thermal emission, we fit a
power law to the spectrum. 
However, the power law fit to this component yields a 
photon index of $\Gamma=0.45$, which is implausible. 

In this spectrum, the column density has decreased 
 by a factor of three; the best fit column density for  $Z=0.3~ Z_\odot$ 
is $\sim3 \times 10^{23}$ cm$^{-2}$. The Fe 6.4 keV line is  not 
present, but the low-temperature component is there. Because of 
few data points and a large uncertainty in the column density in this component, we
again fix the column density to the Galactic value. 
When we fit the low temperature component with a thermal plasma or a 
power law model, it fits with a temperature of $\sim1-2$~keV or a power law
of $\sim1.7$. However, another component  is still required by the data. 
When we try to fit
this component using the same column density as that of the main emission
component, we find $T=0.11$ keV and 
a very high
unabsorbed luminosity  $\sim  10^{45}$ erg s$^{-1}$,  
which is implausible.
Thus we try to fit this component with an 
independently varying column density. The column density associated with this
component is around 1/4 that of the high $N_H$ component column density.
This gives a reasonable and physically plausible component and indicates that the
flux is this component is $15-20$\% of the total emission. 
Thus, in this spectrum, we have three components:
a high-$N_H$ high-$T$ component, a high-$N_H$ low-$T$ component and a low-$N_H$ component.

Although the preferred temperature again reaches the upper bound 
allowed by the Mekal model, the error determination shows that there is no
upper bound and the lower bound to the temperature is 12~keV. 
Our final model thus is
Absorption$\times$Power law $+$
Absorption$\times$Mekal+Absorption$\times$Mekal.
Figure \ref{second} shows the best fit to the
{\em Chandra} October 2011 spectrum.
Table \ref{tab:xspec} lists the models and best fit parameters.
In this case,  81.1\% of the unabsorbed flux is carried by the high column density
component, 18.2\%  by the lower-$N_H$ component and  0.7\% 
 is carried by the power law component.

\section{Results and Interpretation}
\label{discussion}

Here we highlight the main differences between the December 2010 and October 2011
spectra and discuss the best fit models and their implications. 
The lower limits on the temperature for the two spectra are 8~keV and 12~keV,
respectively, showing that a hot component is present.
The column densities of the main X-ray 
emission component (high-$N_H$ component)
are  high at both epochs. The column densities at the first and  second epochs are
$\sim10^{24}$ cm$^{-2}$ and 
$3\times10^{23}$ cm$^{-2}$ (for a metallicity of $Z\approx 0.3 Z_\odot$), respectively.
These are 3000 times and 
1000 times  higher than the Galactic column density ($3\times10^{20}$
cm$^{-2}$). 
The high value and variability of $N_H$ point to an origin in the circumstellar medium.
The excess column density 
to the X-ray emission is not accompanied
by high extinction to the supernova, showing that the column is probably
due to mass loss near the forward shock wave where any dust has been
evaporated.
This is the first time that external
circumstellar X-ray absorption has been clearly
observed in a supernova.

Assuming 2010 Oct 10 as the date of explosion \citep{andrews11,patat11},
the epochs of the two {\em Chandra} observations are 
59 and 373 days, respectively.
The $0.2-10$ keV absorbed flux at the second epoch ($1.1\times10^{-12}$ erg cm$^{-2}$ s$^{-1}$) 
is  higher than that at 
the first epoch ($6.5\times10^{-13}$ erg cm$^{-2}$ s$^{-1}$),
but this is due to the lower absorption
column density at the second epoch. The actual unabsorbed emission from the
SN is  constant within $20-30$\%.
At the two epochs, the unabsorbed luminosity in the $0.2-10$~keV
band is $\sim7\times10^{41}\ergs$, placing
SN 2010jl among the most luminous X-ray supernovae yet observed.
Table 1 of \cite{immler07} shows that the only other SNe with comparable
luminosities are Type IIn events or gamma-ray burst associated SNe at early times.
The  luminosity of $3.6\pm 0.5\times 10^{40}\ergs$ found by {\it Swift} on 2010 Nov 5
\citep{immler10} is revised to a value close to our 
{\it Chandra} result if $N_H\sim 10^{24}$ cm$^{-2}$ is assumed.
In the thermal interpretation, the shock velocity can be deduced
as   $v_{sh}=[16 k T/(3\mu)]^{1/2}=7700(kT/80{\rm~keV})^{1/2}$ km s$^{-1}$, 
where $k$ is Boltzmann's constant
and $\mu$ is the mean particle weight. 
A lower limit of 10 keV for the temperature puts a lower
limit of the 2700~km~s$^{-1}$ on the shock speed.

In comparing the observed luminosity to a thermal emission model to find the physical 
parameters, we note that our measurements give the spectral luminosity, not the
total luminosity.
We use equation (3.11) of \cite{flc96} for the luminosity, adjusting to an observed photon
energy of $\sim 10$ keV rather than $100$ keV; the Gaunt factor is increased to
$2-3$.  
For the preshock column density, we use equation (4.1) of \citet{flc96}.
These expressions allow for a variation of the preshock density $\propto r^{-s}$,
where $s$ is a constant.  
The value $s=2$ corresponds to a steady wind and is commonly used, but implies
stronger evolution than we observe in SN 2010jl.
If the circumstellar medium around SN 2010jl is due to some presupernova eruptive
event, deviation from $s=2$ is plausible.
Another parameter is $m$, determined by the expansion of the supernova shock $R\propto t^m$.
For the plausible value $m=0.8$, we find that $s=1.6$ give a reasonable representation
of the luminosity and $N_H$ evolution.
The implied value of the mass loss rate $\dot M$, normalized to $R=10^{15}$ cm,
is $\dot M_{-3}/v_{w2}\approx 8v_4^{0.6}$, where $\dot M_{-3}=\dot M/(10^{-3}\ml)$,
$v_{w2}$ is the preshock wind velocity in units of $100\kms$,
and $v_4$ is the shock velocity in units of $10^4\kms$ at the first epoch.

The high temperature implies that we are observing the forward shock region.
The physical conditions are such that the forward shock front is close to
the cooling regime \citep{chevalier12}.
In this case, the luminosity of the forward shock is expected to dominate that
from the reverse shock and the reverse shock emission may be absorbed by a cooled shell,
which explains the lack of observational evidence for reverse shock emission.

In modeling the X-ray absorption in SN 2010jl we have assumed that the absorbing gas
is not fully ionized.
If the circumstellar gas is photoionized by the X-ray emission, the absorption is
reduced \citep[e.g.,][]{fransson82}.
Taking an X-ray luminosity of $10^{42}\ergs$ and $\dot M_{-3}/v_{w2}\approx 8$ (at $r=10^{15}$ cm),
the ionization parameter is $\zeta =L/nr^2\approx 200$; a similar value is
obtained taking $nr\sim N_H\sim 10^{24}$ cm$^{-2}$ and $r=6\times 10^{15}$ cm for
the early epoch.
This is in a regime where the CNO elements may be completely ionized, but
Fe is not \citep{hatchett76}.
The CNO elements absorb radiation at $\sim 1$ keV, so there is the possibility
of getting enhanced emission around that energy, as is observed in SN 2010jl.
We investigated this possibility by running various cases with the CLOUDY
photoionization code \citep{fer98}.
It is possible to obtain cases in which there is a peak at $\sim 1$ keV, but they had
too little absorption in the $1.5-3$ keV range.
We thus favor a background source origin for the 1 keV emission, especially
because the emission remains fairly constant over the 2 epochs.

The value of $N_H$ in Dec 2010 implies that $\tau_{es}\approx 1$, where
$\tau_{es}$ is the electron scattering optical depth through the preshock wind.
The H$\alpha$ line at that time showed roughly symmetric broad wings \citep{smith12} that
are probably best explained by electron scattering in the slow moving wind.
\cite{chugai01} estimated that the broad features observed in SN 1998S require
$\tau_{es}\approx (3-4)$.
The required optical depth may be several times that observed along the line of sight to the X-ray emission, which could be the result of asymmetry.
\cite{andrews11} found that the column density of dust needed for observed infrared emission is larger than that on the line of sight to the supernova, although this is at larger radii.

\begin{figure}
\centering
\includegraphics[angle=0,width=0.92\textwidth]{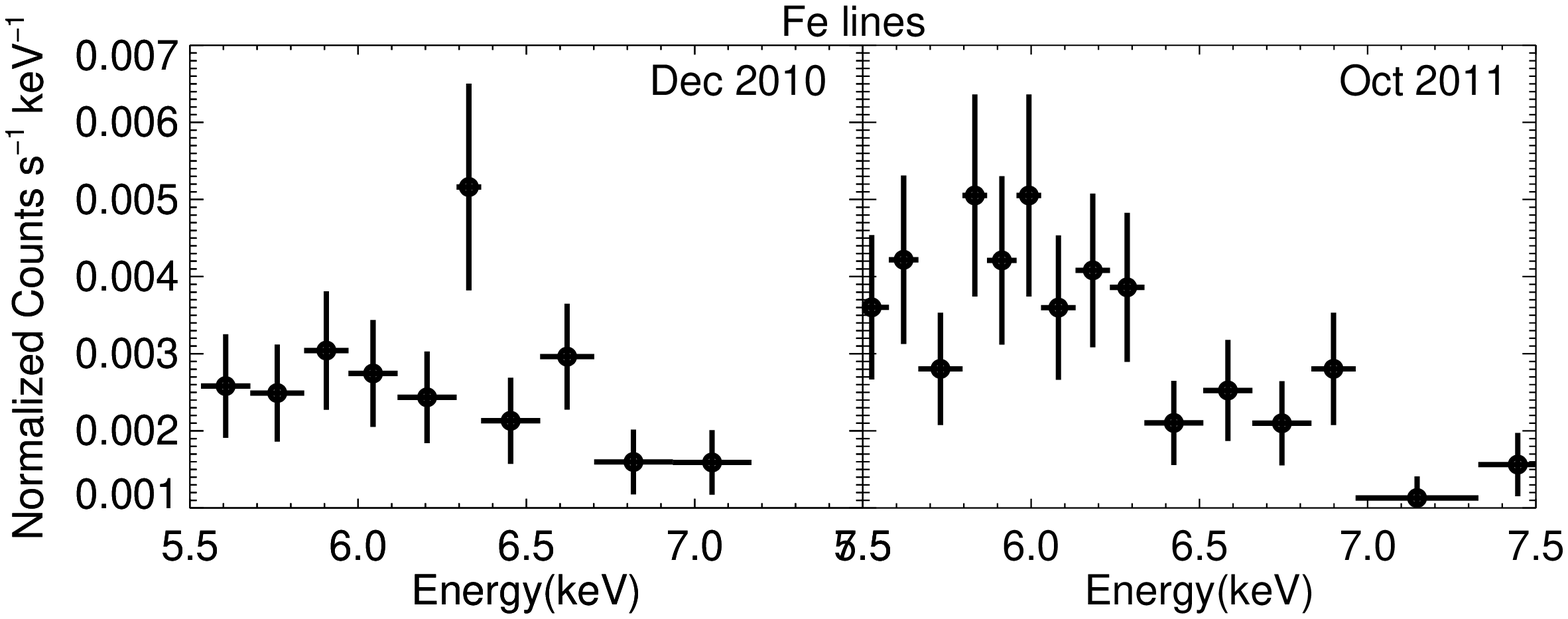}
\caption{Comparison of the December 2010 and October 2011
spectra around the Fe 6.4~keV line energy. The line  is clearly visible
at the early time, but not at the later time.
} 
\label{comparison}
\end{figure}

The Dec 2010 spectrum shows a 6.4 keV feature (Figure \ref{comparison}), which 
 is identified with the narrow K$\alpha$ iron line. Since the
6.4 keV Fe line arises from 
neutral or low ionized iron (Fe\,I to Fe\,XI), it  supports 
our finding that the radiation field is not able to completely ionize the circumstellar gas.
A simple estimate of the expected equivalent width
of the Fe line ($EW_{Fe}$)  can be obtained from
equation (5) of \cite{kallman04}: $EW_{Fe}=0.3(Z/Z_{\odot})N_{24}$ keV, where $N_{24}$ is the
circumstellar column density in units of $10^{24}$ cm$^{-1}$ and the line production is due to
a central X-ray source in a spherical shell.
The expression assumes a flux spectrum $F_E\propto E^{-1}$; $F_E\propto E^{-0.4}$,
which is more appropriate to the hot thermal spectrum here, increases $EW_{Fe}$ by 1.2.
The prediction for the metallicity in our case in the early spectrum is 
$EW_{Fe}=0.1$ keV, and the observed value is 0.2 keV.
In view of the uncertainties in the model and the observations, 
we consider the agreement to be adequate.
At the second epoch, $N_H$ is smaller by a factor of 3, so the strength of the Fe line
should be correspondingly smaller; this is consistent with the nondetection of the line.
The problem with this picture is that it assumes the Fe is in the low ionization stages 
that produce the K$\alpha$ line; this requires an ionization parameter $\zeta \la 5$
\citep{kallman04}, which is below the inferred value.
One possibility is that the circumstellar gas is clumped, with a density $\ga 40$ times the average;
another is that the K line emission is from dense gas that is not along the line of sight.

A thermal fit to the low temperature component implies an absorbing column density
of $(1.37\pm8.44)\times 10^{20}$ cm$^{-2}$, much less than the column 
to the hot component and consistent with the Galactic column density 
within the errors.
This  rules out the possibility that the
cooler X-rays come from slow cloud shocks in the clumpy CSM or from the
reverse shocks. The component is also present in the second epoch.   It
 could arise from a presupernova mass loss event
or from an unrelated source in the direction of the supernova.
The components are best fit with either a thermal component ($T\sim 1-2$ keV) or  a power law
with $\Gamma=1.6-1.7$. 
The luminosities of this component in the
Dec 2010 and Oct 2011 spectra are $3.5\times10^{39}$ erg s$^{-1}$ and
$5\times10^{39}$ erg s$^{-1}$, respectively. 
The luminosity range and the
power law index are compatible with a background ultraluminous X-ray source (ULX),
which can typically be described by an absorbed power law spectrum \citep{swartz04}.
Since the error in the flux determination is between $20-30$\%, a factor
of 1.4 change in the
 luminosity at the two epochs is consistent with a constant flux. 
Thus we attribute this component  to a background source, most likely an ULX, which is associated with 
the blue excess emission region seen in the pre-SN
HST images \citep{smith10}. 
We examined the HEASARC  archives for useful limits on such a source, but did not find any.

The December 2010 spectrum has only one temperature component associated 
with the high column density.
However, in the October 2011 spectrum, there are two temperature components associated
with a high column density, one
with temperature $\ga 10$ keV and another with temperature 1.1 keV.
The lower temperature
component fits with  1/4 the column density of the high temperature
component.
The fact that the component is absent at the first epoch suggests that it is
related to the supernova emission.
We examined the possibility that the emission is the result of reduced
absorption due to photoionization of
the absorbing material, in particular, that lighter atoms are ionized but heavier atoms are not.
However, we were not able to reproduce the observed emission and the source of
this emission remains uncertain.

SN 2010jl is a special Type IIn supernova 
because we have been able to catch it in X-rays  early
on with as sensitive an instrument as {\em Chandra} and trace the
early X-ray evolution.  We observe dramatic changes over two
epochs separated by 10 months.
For the first time we see  clear evidence of external 
CSM absorption in a  supernova. 
We also find that the CSM is not fully photoionized by the SN
emission, the SN is very luminous in X-rays, and the temperature of the emitting gas is $\ga 10$ keV.

\acknowledgments
We are grateful to the referee for useful comments.
Support for this work was provided by  NASA through Chandra Awards  GO0-211080X and  GO2-13082X issued by the Chandra X-ray Observatory Center, which is operated by the Smithsonian Astrophysical Observatory for and on behalf of  NASA under contract NAS8-03060.

{\it Facilities:} \facility{Chandra}.

\clearpage

\begin{deluxetable}{llccccc}
\tabletypesize{\scriptsize}
\tablecaption{Spectral model fits to the SN 2010jl spectra
\label{tab:xspec}}
\tablewidth{0pt}
\tablehead{
\colhead{Spectrum} & \colhead{Model} & \colhead{${\chi}^2/{\nu}$}  &
\colhead{N$_{\rm H}$}  & \colhead{Parameter } & \colhead{Abs. flux} &
\colhead{Unabs. flux}\\
}
\startdata
\multirow{6}{*}{Dec 2010} & {\bf Mekal} & 0.84(23) &
$9.70^{+1.60}_{-1.61} \times10^{23}$
& $kT=79.9^{+\cdots}_{-68.03}$ &
$6.55\times10^{-13}$ & $2.44\times10^{-12}$\\
& ~~+ Gaussian & $\cdots$ & $\cdots$
& $E=6.32^{+0.06}_{-0.06}$  & $2.43\times10^{-14}$ &  $4.25\times10^{-14}$\\
& ~~+ PowerLaw& $\cdots$ & $3.0 \times10^{20}$ (fixed)
& $\Gamma=1.68^{+0.66}_{-0.72}$  & $1.21\times10^{-14}$ &
$1.29\times10^{-14}$\\
& {\bf Mekal} & 0.87(24) & $10.59^{+1.77}_{-1.28} \times10^{23}$
& $kT=8.0$ (fixed) &
$5.57\times10^{-13}$ & $3.23\times10^{-12}$\\
 & ~~+ Gaussian & $\cdots$ & $\cdots$
& $E=6.32^{+0.06}_{-0.06}$  & $3.02\times10^{-14}$ &  $5.56\times10^{-14}$\\
 & ~~+ PowerLaw& $\cdots$ & $3.0 \times10^{20}$ (fixed)
& $\Gamma=1.63^{+0.70}_{-0.64}$  & $1.18\times10^{-14}$ &
$1.24\times10^{-14}$\\ 
\tableline
&  & & & & & \\
\multirow{6}{*}{Oct 2011} & {\bf Mekal} & 0.92(75) &
$2.67^{+3.47}_{-0.48} \times10^{23}$
& $kT=79.9^{+\cdots}_{-68.55}$ &
$1.04\times10^{-12}$ & $2.13\times10^{-12}$\\
& ~~+ Mekal & $\cdots$ & $8.38^{+0.52}_{-0.43} \times10^{22}$
& $kT=1.05^{+0.95}_{-0.44}$  & $3.76\times10^{-14}$ &  $4.94\times10^{-13}$ \\
& ~~+ PowerLaw& $\cdots$ & $3.0 \times10^{20}$ (fixed)
& $\Gamma=1.54^{+0.73}_{-0.71}$  & $2.37\times10^{-14}$ &
$2.42\times10^{-14}$\\
& {\bf Mekal} & 0.99(78) &
$3.51^{+2.52}_{-1.14} \times10^{23}$
& $kT=12$ (fixed) &
$9.49\times10^{-13}$ & $2.62\times10^{-12}$\\
& ~~+ Mekal & $\cdots$ & $9.06^{+3.95}_{-2.42} \times10^{22}$
& $kT=1.15^{+1.41}_{-0.53}$  & $5.34\times10^{-14}$ &  $5.87\times10^{-13}$ \\
& ~~+ PowerLaw& $\cdots$ & $3.0 \times10^{20}$ (fixed)
& $\Gamma=1.76^{+0.81}_{-0.84}$  & $1.97\times10^{-14}$ &
$2.15\times10^{-14}$\\
\enddata
\tablecomments{Here $N_H$ is in cm$^{-2}$, $E$ is in keV,
and the fluxes are in
erg cm$^{-2}$ s$^{-1}$. The fluxes are given in the
$0.2-10.0$~keV energy range. The absorbed and unabsorbed fluxes are
for that particular component in the model.    The
errors in the fluxes are $20-30$\%.}
\end{deluxetable}

\clearpage

\clearpage

\clearpage

\end{document}